\documentclass[aip,apl,reprint]{revtex4-1}
\usepackage{amssymb}
\usepackage{amsmath}
\usepackage{dcolumn}
\usepackage{bm}
\usepackage{graphicx}

\begin{document}

\title{Movable Fiber-Integrated Hybrid Plasmonic Waveguide on Metal Film}
\author{Chang-Ling Zou}
\affiliation{Key Lab of Quantum Information, University of Science
and Technology of China, Hefei 230026, Anhui, P. R. China}
\author{Fang-Wen Sun}
\email{fwsun@ustc.edu.cn} \affiliation{Key Lab of Quantum
Information, University of Science and Technology of China, Hefei
230026, Anhui, P. R. China}
\author{Chun-Hua Dong}
\affiliation{Key Lab of Quantum Information, University of Science
and Technology of China, Hefei 230026, Anhui, P. R. China}
\author{Yun-Feng Xiao}
\affiliation{State Key Lab for Mesoscopic Physics, School of
Physics, Peking University, Beijing 100871, P. R. China}
\author{Xi-Feng Ren}
\affiliation{Key Lab of Quantum Information, University of Science
and Technology of China, Hefei 230026, Anhui, P. R. China}
\author{Liu Lv}
\affiliation{Key Lab of Quantum Information, University of Science
and Technology of China, Hefei 230026, Anhui, P. R. China}
\author{Xiang-Dong Chen}
\affiliation{Key Lab of Quantum Information, University of Science
and Technology of China, Hefei 230026, Anhui, P. R. China}
\author{Jin-Ming Cui}
\affiliation{Key Lab of Quantum Information, University of Science
and Technology of China, Hefei 230026, Anhui, P. R. China}
\author{Zheng-Fu Han}
\affiliation{Key Lab of Quantum Information, University of Science
and Technology of China, Hefei 230026, Anhui, P. R. China}
\author{Guang-Can Guo}
\affiliation{Key Lab of Quantum Information, University of Science
and Technology of China, Hefei 230026, Anhui, P. R. China}

\begin{abstract}
A waveguide structure consisting of a tapered nanofiber on a metal
film is proposed and analyzed to support highly localized hybrid
plasmonic modes. The hybrid plasmonic mode can be efficiently
excited through the in-line tapered fiber based on adiabatic
conversion and collected by the same fiber, which is very convenient
in the experiment. Due to the ultrasmall mode area of plasmonic
mode, the local electromagnetic field is greatly enhanced in this
movable waveguide, which is potential for enhanced coherence light
emitter interactions, such as waveguide quantum electrodynamics,
single emitter spectrum and nonlinear optics.
\end{abstract}

\maketitle

Plasmonics in metal nanostructure are being extensively studied for
its excellent capability of confining light in subwavelength scale
\cite{ozbay,schuller}. Since the local electric field of the
plasmonic mode is dramatically increased at the dielectric-metal
interface, the light-matter interaction can be greatly enhanced.
Thus, over the past few years, metal nanostructures including
nanoparticles, nanowires and nanorings, have been studied for highly
sensitive sensing, surface enhanced Raman scattering, and surface
plasmonic amplification by stimulated emission of radiation (SPASER)
\cite{SPASER,nie,sensor,xiaoprl,zou2,fedutik,chang1,akimov,dong,lin}.
Recently, it has also been found that when an optical emitter (e.g.,
a quantum dot (QD)) is placed around the silver nanowire, its
spontaneous emission can be significantly modified \cite{chang1},
known as the Purcell effect. As a result, the plasmonic modes of
nanowire provide an alternative approach to study the broadband
waveguide quantum electrodynamics (QED)
\cite{chang1,zou2,fedutik,akimov,dong,lin}, and hold great potential
for the single photon source \cite{akimov} and sub-wavelength single
photon transistor \cite{chang2}.

One of the limitation of these metal nanostructures is the high
absorption loss in metal. Very recently, the hybrid dielectric-metal
structures have been proposed, which are consisted of a dielectric
waveguide or resonator near a metal substrate. These hybrid
plasmonic modes can be low loss while remain high electromagnetic
field localization \cite{zhang1,zhang3,xiaoJPB}. In experiments, the
hybrid plasmonic modes are excited and collected through free space.
Unfortunately, this process suffers from low efficiency due to the
mismatching of momentum of light. In this paper, we propose and
numerically investigate a fiber-integrated hybrid plasmonic
waveguide. Based on the adiabatic conversion, the local plasmonic
mode can be excited and collected by fiber with very high
efficiencies. More importantly, the movable structure can be easily
fabricated and controlled, thus offering a great feasibility in
future experiment.

\begin{figure}[tbp]
\centerline{
\includegraphics[width=0.25\paperwidth]{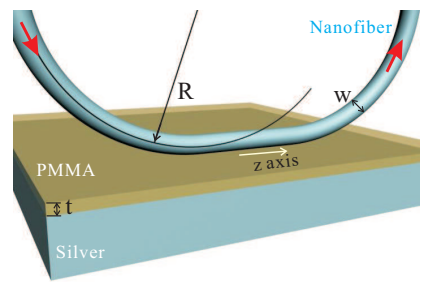}}
\caption{(color online) (a) Schematic diagram of the hybrid
plasmonic waveguide, consisting of a bending nanofiber on a metal
substrate, which is covered by a thin PMMA film (thickness $t$).}
\end{figure}

Fig. 1 shows the geometry of the proposed hybrid system. A curved
silica nanofiber is put on a silver substrate surrounded by air,
with a nano-scale thickness ($t$) PMMA polymer film between them
\cite{PMMA}. The curved nanofiber with a sub-micrometer diameter has
two symmetrical bending parts (radius $R$), which are connected by a
straight part. In the following study, we fix the working wavelength
at $633$ nm. The refractive indices of silica and PMMA are
$n_{\text{silica}}=1.45$ and $n_{\text{PMMA}}=1.49$, repectively. In
experiment, the nanofiber can be pulled from a standard fiber. The
waist of the nanofiber has a uniform diameter ($w$), which can be as
small as $120$ nm \cite{taperexp}. The interaction area of the
nanofiber and substrate can be selected freely by mounting the
nanofiber on a holder system with a high resolution three-axis
translation stage.

At the region where the nanofiber contacts with the substrate, the
hybrid plasmonic modes will appear \cite{zhang1,zhang3,xiaoJPB}. The
heart of our proposal is the adiabatic conversion between the fiber
guiding mode and the plasmonic mode. As the guiding mode propagates
in the bending nanofiber which slowly approaches the metal
substrate, the confined field in dielectric will evolute to the
plasmonic mode at metal-dielectric interface. Inversely, when the
nanofiber slowly deviates from the substrate, the plasmonic mode
will convert back to the guiding mode. The similar adiabatic
conversion of the optical mode from one mode profile to the other
has also been extensively studied in fiber and waveguide optics \cite%
{adiabatic1,adiabatic2}. For example, in the slowly varying fiber
taper in our structure, the fundamental single-mode fiber mode can
be adiabatically converted to the fundamental nanofiber mode with a
large evanescent tails extended to the air, with the loss less than
$0.01$. Therefore, the plasmonic mode can be excited and collected
efficiently through a standard single-mode fiber.

Now, we analyze the adiabatic process in the present structure. The
light propagates along the waveguide, defined as z-axis. Similar to
the time evolution of quantum states in Schrodinger function, the
evolution of light in waveguides can be expressed as
\cite{adiabatic1,adiabatic2}

\begin{equation}
i\frac{\partial }{\partial z}\left\vert \varphi \left( z\right)
\right\rangle =H(z)\left\vert \varphi \left( z\right) \right\rangle
, \label{1}
\end{equation}%
where $H(z)$ is position (z) dependent Hamiltonian due to the
variation of waveguide-substrate space, and $\left\vert \varphi
\left( z\right) \right\rangle $ is the wavefunction at the cross
section. At any position of the hybrid waveguide, there exist
instantaneous nondegenerate eigenstates of $H(z)$. In the basis of
eigenmodes $\{\beta _{m},\left\vert \varphi _{m}\left( z\right)
\right\rangle \}$, the electromagnetic field can be expanded as

\begin{equation}
\left\vert \varphi \left( z\right) \right\rangle =\sum
a_{m}(z)e^{-i\int \beta _{m}(z)dz}\left\vert \varphi _{m}\left(
z\right) \right\rangle ,\label{2}
\end{equation}
where $\beta _{m}$ is the propagation constant of mode $\left\vert
\varphi _{m}\left( z\right) \right\rangle $, and
$a_{m}(z)=\left\langle \varphi _{m}\left( z\right) \right\vert
\varphi \left( z\right) \rangle $ is the corresponding coefficient.
Substituting Eq.(\ref{1}) and Eq.(\ref{2}) into the
expression $\left\langle \varphi _{k}\left( z\right) \right\vert i\frac{%
\partial }{\partial z}\left\vert \varphi \left( z\right) \right\rangle $, we
can obtain
\begin{equation}
\frac{\partial }{\partial z}a_{k}(z)=-\sum g_{km}(z)e^{-i\int (\beta
_{m}(z)-\beta _{k}(z))dz}a_{m}(z),  \label{3}
\end{equation}%
where $g_{km}(z)=\left\langle \varphi _{k}\left( z\right) \right\vert \frac{%
\partial }{\partial z}\left\vert \varphi _{m}\left( z\right) \right\rangle $%
. By solving the equation $\left\langle \varphi _{k}\left( z\right)
\right\vert \frac{\partial }{\partial z}(H(z)\left\vert \varphi
_{m}\left( z\right) \right\rangle )=$ $\left\langle \varphi
_{k}\left( z\right) \right\vert \frac{\partial }{\partial
z}(\beta_{m} (z)\left\vert \varphi _{m}\left( z\right) \right\rangle
)$, we obtain

\begin{equation}
g_{km}(z)=\frac{\left\langle \varphi _{k}\left( z\right) \right\vert \frac{%
\partial H(z)}{\partial z}\left\vert \varphi _{m}\left( z\right)
\right\rangle }{\beta _{k}(z)-\beta _{m}(z)},  \label{4}
\end{equation}%
when $k\neq m$. Therefore, we can finally obtain

\begin{eqnarray}
\frac{\partial }{\partial z}a_{k}(z) &=&-\left\langle \varphi _{k}\left(
z\right) \right\vert \frac{\partial }{\partial z}\left\vert \varphi
_{k}\left( z\right) \right\rangle a_{k}(z)  \notag \\
&&-\sum_{n\neq m}g_{km}(z)e^{i\int (\beta _{m}(z)-\beta _{n}(z))dz}a_{m}(z).
\label{5}
\end{eqnarray}

The evolution of states in the hybrid waveguide at the bending
region can be obtained by solving the array of equations for all
$a_{k}(z)$. The first term of Eq.(\ref{5}) stands for the Berry
phase, which does not play a significant role here since we just
concern about the energy conversion. The second term represents the
coupling between eigenmodes, which reveals the requirement for
efficiently adiabatic mode conversion, i.e. $g_{km}(z)\ll 1$. Since
the variance of $H(z)$ depends on the air space ($s$) between
waveguide and substrate, we can rewrite $g_{km}(z)$ as
$g_{km}(s)\frac{\partial s}{\partial z}$ where
$g_{km}(s)=\left\langle \varphi _{k}\left( s\right) \right\vert
\frac{\partial H(s)}{\partial s}\left\vert \varphi _{m}\left(
s\right) \right\rangle /(\beta _{k}(s)-\beta _{m}(s))$. When
nanofiber approaches the substrate ($s \approx \lambda$),
$\frac{\partial s}{\partial z}\approx -\sqrt{\frac{s}{2R}}$.
Therefore, the adiabatic conversion requires $R\gg \lambda $. In
addition, the nanofiber should work in the single mode regime to
exclude high order modes, with the single mode criterion $V=(\pi
w/\lambda )\sqrt{n^{2}-1}<2.405$.

\begin{figure}[tbp]
\centerline{
\includegraphics[width=0.3\paperwidth]{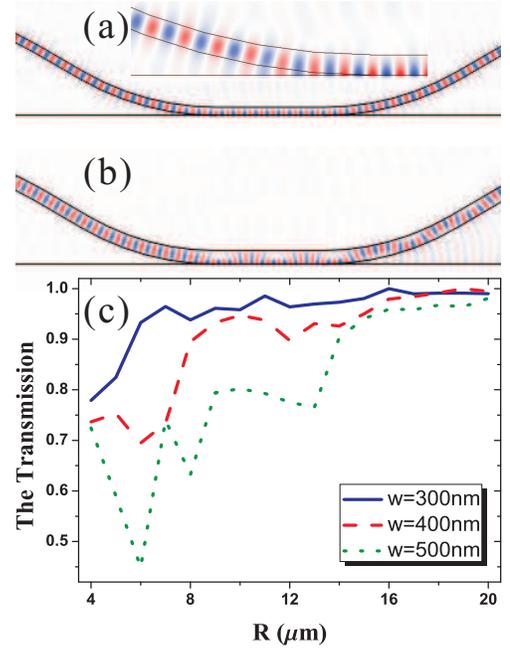}}
\caption{(color online) (a) and (b) is the filed distributions for
$w=300\mathrm{nm}$ and $w=500\mathrm{nm}$ waveguide respectively,
with $R=10\mathrm{\mu m}$ and $t=0$.(c) The transmission of the
hybrid waveguide for different $w$ and $R$, with $t=0$. Inset: the
detail of adiabatic conversion}
\end{figure}

In order to verify the losslessly conversion of the energy between
fundamental modes under reasonable parameters, we numerically solve
the stationary harmonic propagation of optical field in the
nanofiber contacting with silver by two-dimensional finite element
method. Due to the complex boundary condition in our structure, we
can not analytically solve it by the coupled-mode equation Eq.(5).
Figs. 2(a) and 2(b) shows the field distribution as light
propagating in the hybrid waveguide. For $w=300 \mathrm{nm}$, the
fundamental dielectric modes are almost totally converted to
plasmonic mode, as shown in the inset of Fig. 2(a). However, for
$w=500 \mathrm{nm}$, we can find that the mode is not totally
converted to the plasmonic mode, and the mode in the contact region
and transmitted light is multimode. The metal loss is neglected
here, so the transmission directly indicates the adiabatic
conversion efficiency. Fig. 2(c) shows the transmission of the
hybrid waveguide structure, where the waveguide mode converts to the
plasmonic mode and then converts back. Clearly, when the radius $R$
is larger, the conversion efficiency is higher. When $R$ is small,
the dynamics is complex, since the adiabatic condition is not
fulfilled. For different nanofiber diameter $w$ (Fig. 1(c)), thin
fiber shows better preformation. The dependence of transmission on
$w$ and $R$ is agree with our theoretical analysis of adiabatic
conversion.

As the lossless adiabatic conversion is confirmed, we turn to
analyze the properties of the hybrid plasmonic mode. For reasonable
parameters, with $R=20$ $\mu $\textrm{m}, and $w=300$ nm, the
conversion efficiency is about $99\%$. In the contact region, the
cross section is uniform along z-axis, and the hybrid plasmonic mode
propagates harmonically along the waveguide. We can investigate the
energy distribution at the two-dimensional cross section\cite{zou},
as shown in Figs. 3(a) and 3(b). For the transverse electric (TE)
polarization, the energy is localized at the nanofiber-substrate
interface, while the transverse magnetic (TM) mode is well confined
in the nanofiber. It is evident that the mode area of TE mode is
greatly reduced, and the maximum of the electric field is located
around nanofiber-substrate interface. We plot the mode area ($A$)
and propagation length ($L$) \cite{AL} against the waveguide width
($w$) for different $t$ in Figs. 3(c) and 3(d).

\begin{figure}[tbp]
\centerline{
\includegraphics[width=0.35\paperwidth]{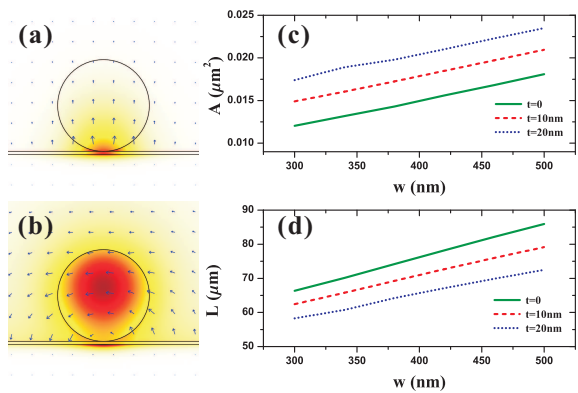}}
\caption{(color online) The energy density distribution at the cross
section of the hybrid waveguide for the TE mode (a) and the TM modes
(b), with $w=300\mathrm{nm}$ and $t=10\mathrm{nm}$. The arrow
indicate the direction of electric field. Mode area (c) and
propagation length (d) of the hybrid plasmonic mode (a) with
$t=0\mathrm{nm}, 10\mathrm{nm}, 20\mathrm{nm}$ for various waveguide
width.}
\end{figure}

Due to the one-dimensional waveguide confinement, the density of
states changes near the nanofiber. As a result, the spontaneous
emission rate is modified, which is known as Purcell effect. The
maximum emission enhancement can be expressed as \cite{xiao}

\begin{equation}
F_{p}=\frac{3n_{eff}}{4\pi }\frac{(\lambda /n)^{2}}{A}.
\end{equation}
For nanofiber with $w=300nm$, and $g=10nm$, we get the enhancement
of spontaneous emission rate are $F_{p}=8.5$ for a emitter in air
and $F_{p}=5.2$ for emitter embed in PMMA. The corresponding
collection efficiency of the single emitter fluorescence is
$F_{p}/(1+F_{p})=88.2\%$ in air and $80.1\%$ in PMMA. Combining with
high efficiency adiabatic conversion, we can finally obtain a high
collection efficiency directly by fiber.

By a 3D translation stage, this hybrid waveguide can be manipulated
or moved easily. Benefiting from the greatly enhanced light-matter
interaction, this hybrid plasmonic waveguide structure is potential
for experimental realization of the single photon transistors
\cite{chang2,shen} or phase flip gate for quantum information
science. It should be noted that the enhancement of light emitter
interaction is broadband, permitting efficient interaction with
quantum dots which have broad emission spectra. For single emitters
disperse on the substrate, this hybrid structure can also be use for
single photon source or spectrum analyze \cite{kartik}. Further
improvement on this structure to enhance the emitter-light
interaction can be done by introducing the bragg-grating mirrors on
the nanofiber \cite{hakuta}.

In conclusion, we have proposed and numerically studied the
adiabatic conversion of energy between a dielectric nanofiber and a
nanofiber-metal film hybrid plasmonic waveguide structure. The
conversion efficiency can exceed $99\%$, and the movable structure
is convenient to be realized in experiment. The hybrid plasmonic
mode has very small mode area and low optical loss, permitting
strong coherence light-matter interaction, which is potential for
studying broadband waveguide QED, single emitter spectrum with high
collection efficiency, and nonlinear optics.

The work was supported by the National Fundamental Research Program of China
under Grant No. 2011CB921200, the Knowledge Innovation Project of Chinese
Academy of Sciences, National Natural Science Foundation of China under
Grant No.11004184 and No.10904137.


\begin{thebibliography}{99}
\bibitem{ozbay} E. Ozbay, Science \textbf{311}, 189 (2006).

\bibitem{schuller} J. A. Schuller et al., Nature Materials \textbf{9},
193-204 (2010).

\bibitem{sensor} J. M. Bingham et al., J. Am. Chem. Soc. \textbf{132}, 17358
(2010).

\bibitem{xiaoprl} Y.-F. Xiao et al., Phys. Rev. Lett. \textbf{105}, 153902
(2010).

\bibitem{nie} S. Nie et al., Science \textbf{275} 1102 (1997).

\bibitem{SPASER} M. A. Noginov et al., Nature \textbf{460},1110 (2009)

\bibitem{chang1} D. E. Chang et al., Phys. Rev. Lett. \textbf{97}, 053002
(2006).

\bibitem{fedutik} Y. Fedutik et al., Phys. Rev. Lett. \textbf{99},
136802(2007).

\bibitem{akimov} A. V. Akimov et al., Nature \textbf{450}, 402 (2007).

\bibitem{zou2} C.-L. Zou et al., J. Opt. Soc. Am. B \textbf{27}, 2495(2010).

\bibitem{dong} C.-H. Dong et al., Appl. Phys. Lett. \textbf{95},
221009(2009).

\bibitem{lin} Z.-R. Lin et al., Phy. Rev. B \textbf{82}, 241401(R) (2010).

\bibitem{chang2} D. E. Chang et al., Nature Phys. \textbf{3}, 807 (2007).

\bibitem{zhang1} R. F. Oulton et al., Nature Photon. \textbf{2}, 496 (2008).

\bibitem{zhang3} R. F. Oulton et al., Nature \textbf{461}, 629 (2009).

\bibitem{xiaoJPB} Y.-F. Xiao et al., J. Phys. B \textbf{43}, 035402 (2010).

\bibitem{PMMA} In experiments, emitters can be dispersed into PMMA and fixed by the nanofilm.

\bibitem{taperexp} A. Stiebeiner et al., Opt. Express \textbf{18}, 22677
(2010).

\bibitem{adiabatic1} M. Skorobogatiy et al., Opt. Express \textbf{10}, 1227
(2002).

\bibitem{adiabatic2} S. G. Johnson et al., Phys. Rev. E 66, 066608 (2002).

\bibitem{zou} C.-L. Zou et al., Appl. Phys. Lett. \textbf{97}, 183102 (2010).

\bibitem{AL} The defination of A and L can be found in ref.\cite{zou}.

\bibitem{xiao} Y.-F. Xiao et al., arXiv: 1010.5067.

\bibitem{shen} J. T. Shen, and S. Fan, Opt. Lett. \textbf{30}, 15 (2005).


\bibitem{kartik} M. I. Davanco and K. Srinivasan, Opt. Lett. \textbf{34},
2542 (2009).

\bibitem{hakuta} F. L. Kien and K. Hakuta, Phys. Rev. A \textbf{81}, 023812
(2010).


\end{thebibliography}
\end{document}